\documentclass[conference,10pt,letterpaper]{IEEEtran}%
\usepackage{amsmath}
\usepackage{graphicx}
\usepackage{multirow}
\usepackage[none]{hyphenat}
\usepackage{float}
\usepackage{subfig}
\usepackage{dblfloatfix}
\usepackage{t1enc}
\usepackage{times}

\usepackage{cite}
\usepackage{amsmath,amssymb,amsfonts}
\usepackage{algorithmic}
\usepackage{fancybox}
\usepackage{pdfpages}
\usepackage{comment}
\usepackage{caption}
\usepackage{graphicx}
\usepackage{mathtools, cuted}
\usepackage{lipsum, color}
\usepackage{comment}
\DeclareUnicodeCharacter{00A0}{~}
\usepackage{textcomp}
\usepackage{xcolor}
\def\BibTeX{{\rm B\kern-.05em{\sc i\kern-.025em b}\kern-.08em
    T\kern-.1667em\lower.7ex\hbox{E}\kern-.125emX}}

\makeatletter

\def\@IMSauthorblockNAMEstyle{\normalfont\IMSauthorsize}
\def\@IMSauthorblockAFFILstyle{\normalfont\IMSaffilsize}
\def\@IMSauthorblockEMAILstyle{\normalfont\IMSaffilsize}
\def\IMSauthorblockNAME#1{%
\relax\@IMSauthorblockNAMEstyle%
#1%
}%
\def\IMSauthorblockAFFIL#1{%
\relax\@IMSauthorblockAFFILstyle%
\vskip\@IEEEauthorblockAtopspace
#1%
}%
\def\IMSauthorblockEMAIL#1{%
\relax\@IMSauthorblockEMAILstyle%
\vskip\@IEEEauthorblockAtopspace
#1%
}%
\newcommand{\IMSauthor}[1]{%
\ifIsBlindReviewVersion%
\author{\phantom{\parbox{\textwidth}{\center\relax#1}}}%
\else%
\author{\parbox{\textwidth}{\center\relax#1}}%
\fi%
}%
\newif\ifIsBlindReviewVersion

\def\IMSthispaperforfinalpublication{\IsBlindReviewVersionfalse}
\def\@maketitle{\newpage
\bgroup\par\addvspace{0.5\baselineskip}\centering%
\ifCLASSOPTIONtechnote
   {\bfseries\large\@IEEEcompsoconly{\sffamily}\@title\par}\vskip 1.3em{\lineskip .5em\@IEEEcompsoconly{\sffamily}\@author
   \@IEEEspecialpapernotice\par{\@IEEEcompsoconly{\vskip 1.5em\relax
   \@IEEEtitleabstractindextextbox{\@IEEEtitleabstractindextext}\par
   \hfill\@IEEEcompsocdiamondline\hfill\hbox{}\par}}}\relax
\else
   \vskip0.2em{\IMStitlesize\ifCLASSOPTIONtransmag\bfseries\LARGE\fi\@IEEEcompsoconly{\sffamily}\@IEEEcompsocconfonly{\normalfont\normalsize\vskip 2\@IEEEnormalsizeunitybaselineskip
   \bfseries\Large}\@title\par}\vskip1.0em\par
   \ifCLASSOPTIONconference%
      {\@IEEEspecialpapernotice\mbox{}\vskip\@IEEEauthorblockconfadjspace%
       \mbox{}\hfill\begin{@IEEEauthorhalign}\@author\end{@IEEEauthorhalign}\hfill\mbox{}\par}\relax
   \else
      \ifCLASSOPTIONpeerreviewca
         {\@IEEEcompsoconly{\sffamily}\@IEEEspecialpapernotice\mbox{}\vskip\@IEEEauthorblockconfadjspace%
          \mbox{}\hfill\begin{@IEEEauthorhalign}\@author\end{@IEEEauthorhalign}\hfill\mbox{}\par
          {\@IEEEcompsoconly{\vskip 1.5em\relax
           \@IEEEtitleabstractindextextbox{\@IEEEtitleabstractindextext}\par\hfill
           \@IEEEcompsocdiamondline\hfill\hbox{}\par}}}\relax
      \else
         \ifCLASSOPTIONtransmag
           {\@IEEEspecialpapernotice\mbox{}\vskip\@IEEEauthorblockconfadjspace%
            \mbox{}\hfill\begin{@IEEEauthorhalign}\@author\end{@IEEEauthorhalign}\hfill\mbox{}\par
           {\vspace{0.5\baselineskip}\relax\@IEEEtitleabstractindextextbox{\@IEEEtitleabstractindextext}\vspace{-1\baselineskip}\par}}\relax
         \else
           {\lineskip.5em\@IEEEcompsoconly{\sffamily}\sublargesize\@author\@IEEEspecialpapernotice\par
           {\@IEEEcompsoconly{\vskip 1.5em\relax
            \@IEEEtitleabstractindextextbox{\@IEEEtitleabstractindextext}\par\hfill
            \@IEEEcompsocdiamondline\hfill\hbox{}\par}}}\relax
         \fi
      \fi
   \fi
\fi\par\addvspace{0.0\baselineskip}\egroup}

\def\IMStitlesize{\@setfontsize{\IMStitlesize}{18}{21pt}}
\def\IMSauthorsize{\@setfontsize{\IMSauthorsize}{12}{13pt}}
\def\IMSaffilsize{\@setfontsize{\IMSaffilsize}{12}{13pt}}
\def\IMScaptionsize{\@setfontsize{\IMScaptionsize}{8}{9pt}}
\def\IMSbibsize{\@setfontsize{\IMSbibsize}{8}{9pt}}

\def\@IEEEauthorblockNstyle{\IMSauthorsize\@IEEEcompsocnotconfonly{\sffamily}\@IEEEcompsocconfonly{\large}}
\def\@IEEEauthorblockAstyle{\IMSaffilsize\@IEEEcompsocnotconfonly{\sffamily}\@IEEEcompsocconfonly{\itshape}\@IEEEcompsocconfonly{\large}}
\def\@IEEEauthordefaulttextstyle{\IMSauthorsize\@IEEEcompsocnotconfonly{\sffamily}\sublargesize}

\def\thebibliography#1{\section*{\refname}%
    \addcontentsline{toc}{section}{\refname}%
    \IMSbibsize\@IEEEcompsocconfonly{\small}\vskip 0.3\baselineskip plus 0.1\baselineskip minus 0.1\baselineskip
    \list{\@biblabel{\@arabic\c@enumiv}}%
    {\settowidth\labelwidth{\@biblabel{#1}}%
    \leftmargin\labelwidth
    \advance\leftmargin\labelsep\relax
    \itemsep \IEEEbibitemsep\relax
    \usecounter{enumiv}%
    \let\p@enumiv\@empty
    \renewcommand\theenumiv{\@arabic\c@enumiv}}%
    \let\@IEEElatexbibitem\bibitem%
    \def\bibitem{\@IEEEbibitemprefix\@IEEElatexbibitem}%
\def\newblock{\hskip .11em plus .33em minus .07em}%
\ifCLASSOPTIONtechnote\sloppy\clubpenalty4000\widowpenalty4000\interlinepenalty100%
\else\sloppy\clubpenalty4000\widowpenalty4000\interlinepenalty500\fi%
    \sfcode`\.=1000\relax}

\long\def\@makecaption#1#2{%
\ifx\@captype\@IEEEtablestring%
\par\@IEEEtabletopskipstrut
\else
\@IEEEfigurecaptionsepspace
\fi
\setbox\@tempboxa\hbox{\normalfont\IMScaptionsize {#1.}\nobreakspace\nobreakspace #2}%
\ifdim \wd\@tempboxa >\hsize%
\setbox\@tempboxa\hbox{\normalfont\IMScaptionsize {#1.}\nobreakspace\nobreakspace}%
\parbox[t]{\hsize}{\normalfont\IMScaptionsize\noindent\unhbox\@tempboxa#2}%
\else
\ifCLASSOPTIONconference \hbox to\hsize{\normalfont\IMScaptionsize\hfil\box\@tempboxa\hfil}%
\else \hbox to\hsize{\normalfont\IMScaptionsize\box\@tempboxa\hfil}%
\fi\fi
\ifx\@captype\@IEEEtablestring%
\@IEEEtablecaptionsepspace
\else
\fi}

\newlength\tablecaptiontotableskip
\newlength\figuretocaptionskip
\def\@IEEEfigurecaptionsepspace{\vskip\figuretocaptionskip\relax}%
\def\@IEEEtablecaptionsepspace{\vskip\tablecaptiontotableskip\relax}%

\def\abstract{\normalfont%
\@IEEEabskeysecsize\bfseries\textit{\abstractname}\,\bfseries\textit{---}\,%
\@IEEEgobbleleadPARNLSP}%

\def\IEEEkeywords{\normalfont%
\@IEEEabskeysecsize\bfseries\textit{\IEEEkeywordsname}\,\bfseries\textit{---}\,%
\@IEEEgobbleleadPARNLSP}%
\def\endIEEEkeywords{\relax\vspace{0.67ex}%
\par\if@twocolumn\else\endquotation\fi%
\normalsize\normalfont}%

\def\@IEEEauthorblockNtopspace{0ex}
\def\@IEEEauthorblockAtopspace{1mm}
\def\IEEEkeywordsname{Keywords}
\def\subsubsection{\@startsection{subsubsection}{3}{\z@}{1.5ex plus 1.5ex minus 0.5ex}%
{0.7ex plus .5ex minus 0ex}{\normalfont\normalsize\itshape}}%
\def\@seccntformat#1{\csname the#1dis\endcsname\relax}
\def\thesubsectiondis{{\hbox to\parindent{\Alph{subsection}.}}}
\def\thesubsubsectiondis{{\hbox to \parindent{\arabic{subsubsection})}}}
\def\theparagraphdis{{\hbox to \parindent{\alph{paragraph})}}}
\IEEEilabelindentA \parindent
\IEEEilabelindent \IEEEilabelindentA
\IEEEelabelindent \parindent
\IEEEdlabelindent \parindent
\IEEElabelindent \parindent

\newlength\@IMSparindent
\newcommand\IMSdisplayacksection[1]{%
\ifIsBlindReviewVersion%
\noindent\phantom{\parbox[t]{\columnwidth}{\normalbaselines\setlength{\parindent}{\@IMSparindent}{#1}\strut}}
\else%
\noindent\parbox[t]{\columnwidth}{\normalbaselines\setlength{\parindent}{\@IMSparindent}{#1}\strut}%
\fi%
}%

\makeatother

\begin{document}
\raggedbottom
%
%
%

\title{Channel Modeling for Physically Secure Electro-Quasistatic In-Body to Out-of-Body Communication with Galvanic Tx and Multimodal Rx}

%
%
%

\IMSthispaperforfinalpublication

\IMSauthor{%
\IMSauthorblockNAME{
Arunashish Datta, Mayukh Nath, Baibhab Chatterjee, Nirmoy Modak, Shreyas Sen,}
\\%
\IMSauthorblockAFFIL{School of ECE, Purdue University, West Lafayette, USA}
\\%
\IMSauthorblockEMAIL{ \{datta30, nathm, bchatte, nmodak, shreyas\}@purdue.edu}

}

%
\maketitle
%
%
%
\begin{abstract}
Increasing number of devices being used in and around the human body has resulted in the exploration of the human body as a communication medium. In this paper, we design a channel model for implantable devices communicating outside the body using physically secure Electro-Quasistatic Human Body Communication. A galvanic receiver shows $5 dB$ lower path loss than capacitive receiver when placed close to transmitter whereas a capacitive receiver has around $15 dB$ lower path loss for larger separation between the transmitter and receiver. Finite Element Method (FEM) based simulations are used to analyze the communication channel for different receiver topologies and experimental data is used to validate the simulation results.

\end{abstract}
%
%

\section{Introduction}

Implantable devices like pacemakers and insulin pumps have become a part of the livelihood of millions of people around the world. These devices coupled with remote patient monitoring has transformed the treatment procedure. However, these implantable devices typically use radio frequency (RF) based wireless protocols which being radiative, leaks critical information outside the body which can potentially be used by expert hackers resulting in fatal consequences. These security concerns have been highlighted by recent US Department of Homeland Security advisories \cite{FDA}. Human Body Communication (HBC) \cite{sen2020body,Health_Monitoring , ISLPED, SocialHBC, datta2020advanced, modak2020bio} has come up as a physically secure \cite{NSR,nath2021inter, maity2021sub,maity2020415} and energy efficient \cite{JSSC} alternative to RF based communication. HBC uses the conductive properties of human tissues to communicate between devices around the body. \par

\begin{figure}[h!]
\centering
\includegraphics[width=0.4\textwidth]{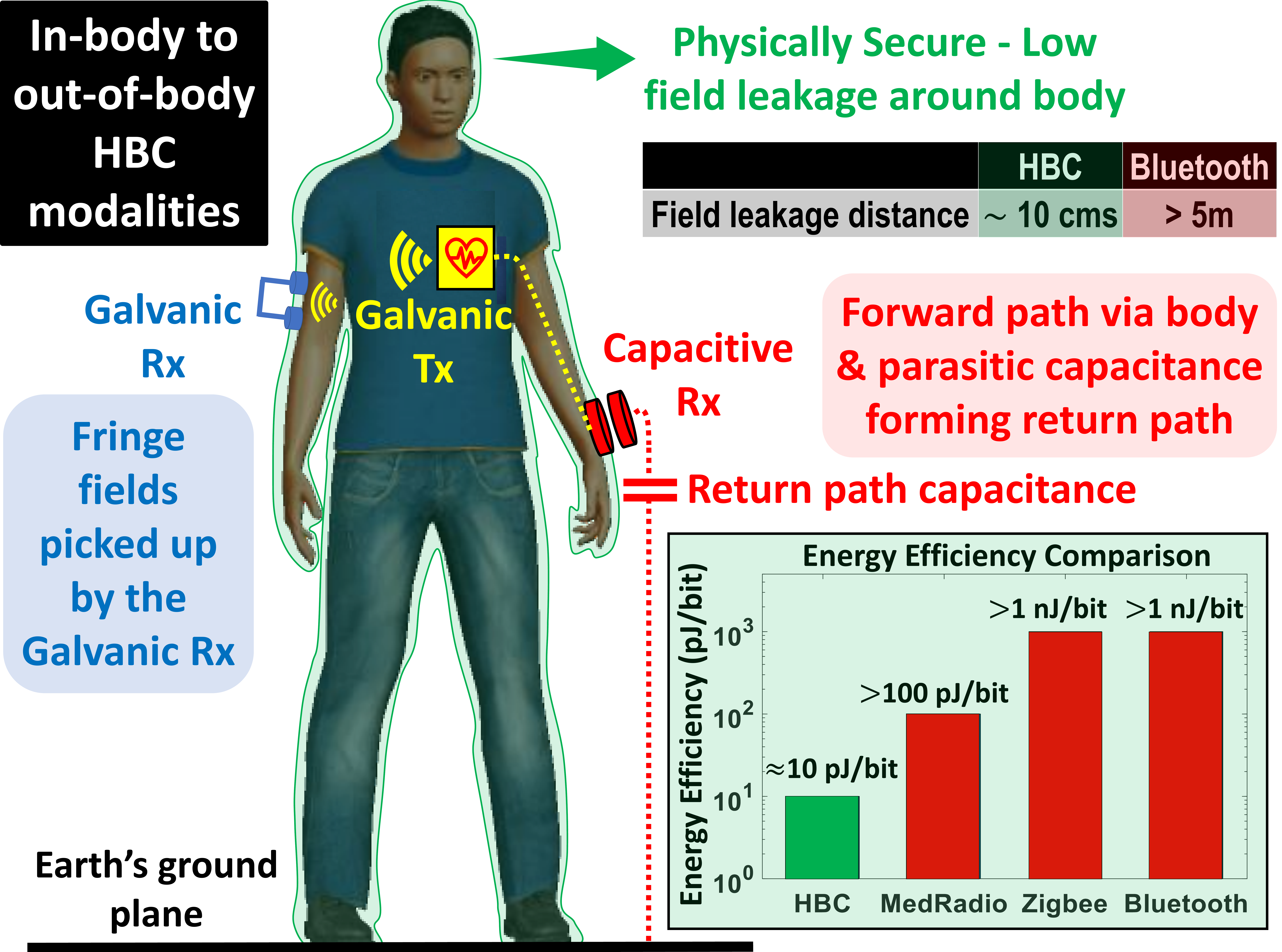}
\caption{HBC modalities for in-body to out-of-body communication}
\vspace{-1.5em}
\label{fig:Intro}
\end{figure}

HBC has two modalities: Capacitive and Galvanic (Fig. \ref{fig:Intro}). In Capacitive HBC \cite{zimmerman1996personal}, a single electrode is used to couple the signal to the body forming the forward path and the parasitic capacitance from earth's ground to floating ground plate of the device forms the return path. In Galvanic HBC \cite{wegmueller2009signal}, differential signal is applied across a pair of electrodes and the fringe fields passing through the body are picked up by the receiver. In this study, we analyze the different modes of Electro-Quasistatic (EQS) HBC \cite{maity2018bio} and design a channel model for in-body to out-of-body communication which can be used as a basis for the development of physically secure and energy efficient implantable communication systems using HBC. 

\begin{figure}[b!]
\vspace{-1.5em}
\centering
\includegraphics[width=0.5\textwidth]{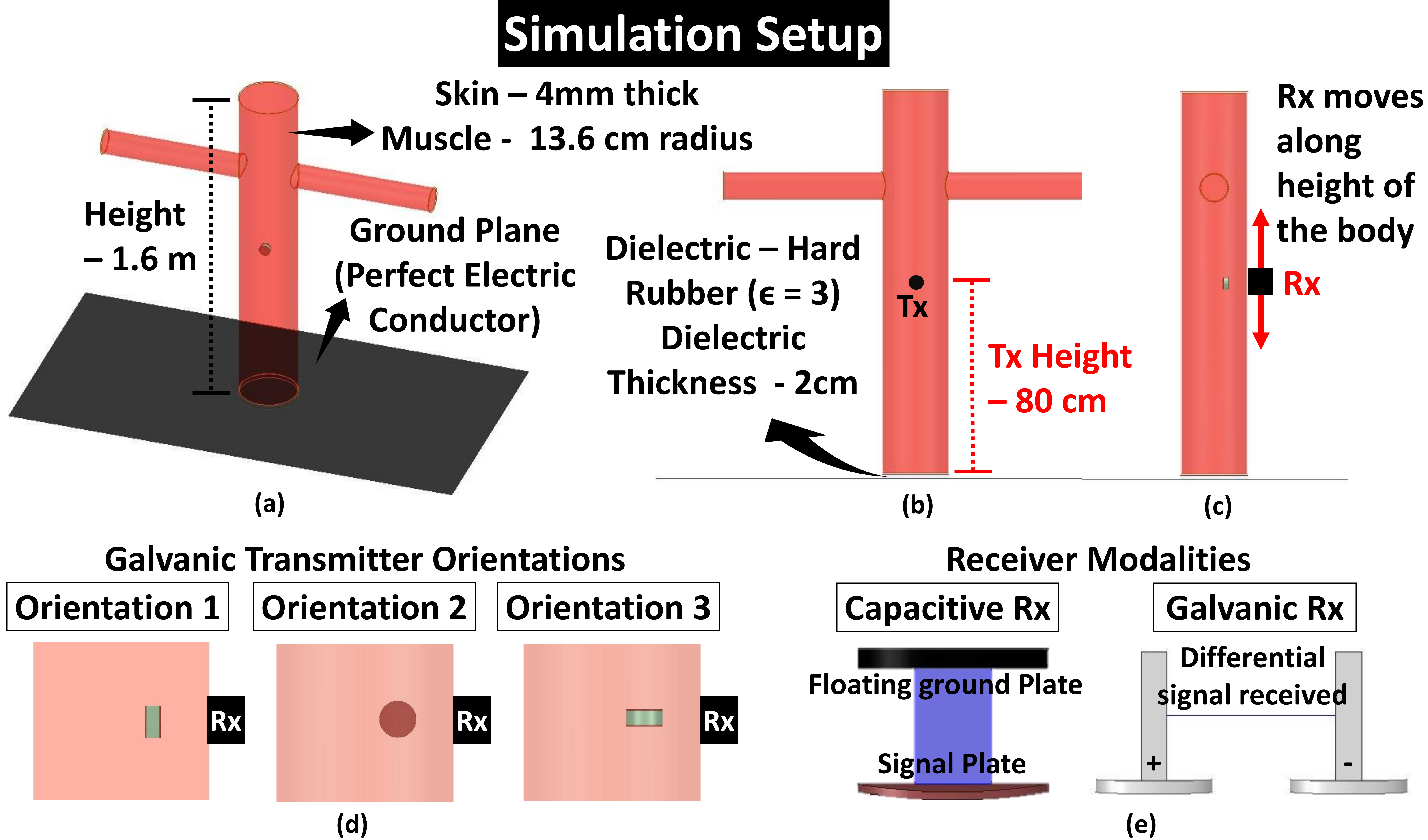}
\caption{(a) Simplified human model used for HFSS simulations and its dimensions. (b)Front view and (c) Side view of the model with the position of Tx and Rx shown. (d) The Tx placed in 3 different orientations shown zoomed in from the side view. (e) Capacitive and Galvanic modes of Rx used for simulations. }
\vspace{-1.5em}
\label{fig:Sim_Setup}
\end{figure}
\begin{figure*}[t!]
\centering
\includegraphics[width=0.85\textwidth]{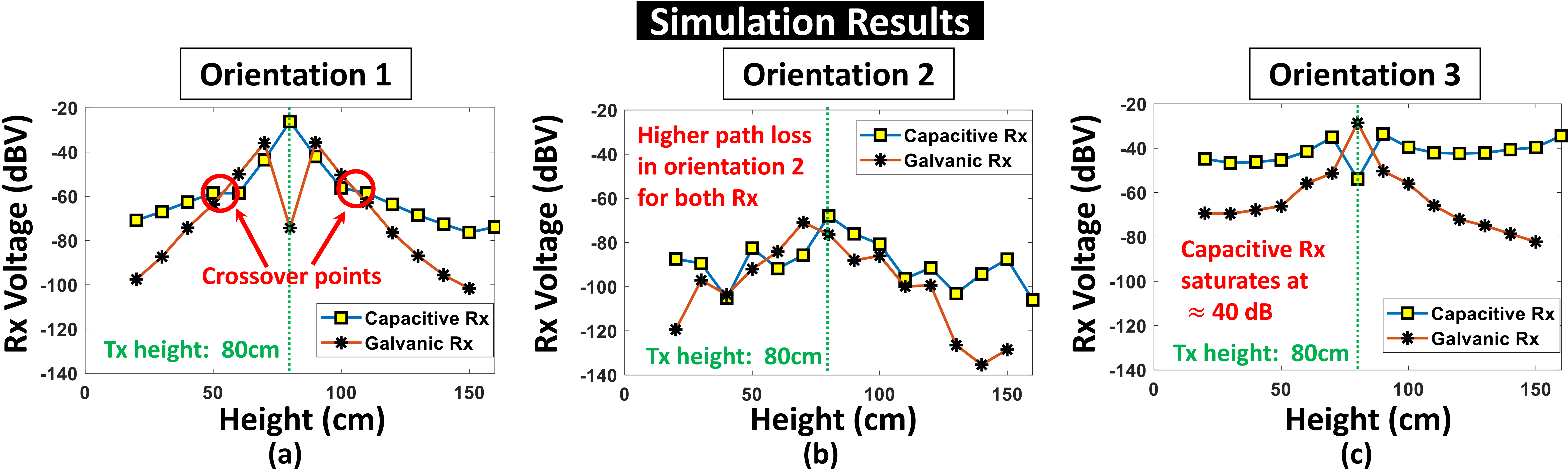}
\vspace{-0.5em}
\caption{Results of the simulations for a capacitive receiver and for a galvanic receiver is compared for (a) Orientation 1, (b) Orientation 2 and (c) Orientation 3.}
\vspace{-1.5em}
\label{fig:Sim_Results}
\end{figure*}
\section{In-Body to Out-of-Body Communication }

\subsection{Physical Layer Security in EQS HBC }
EQS HBC has been shown to be physically secure by operating at low frequencies of upto few $10$s of MHz. This increases the wavelength of the transmitted signals $(>30m)$ making them an order of magnitude more than the dimensions of the human body. Thus, the human body behaves as an electrically small and inefficient antenna which doesn't radiate the signals well. This ensures that the signal remains contained in the body and thus ensures physical layer security for EQS HBC. Signal leakage from EQS HBC has been observed to be of less than $1 cm$ away from body and less than $15cm$ away from the device as compared to more than $5 m$ for RF based protocols like Bluetooth. In this study, for the first time in literature, we explore EQS HBC systems for in-body to out-of-body communication.

For implantable applications, the transmitter (Tx) being inside the body has no direct path to earth's ground making the return path capacitance zero. Hence, a capacitive Tx can not be used for in-body to out-of-body communication and the Tx for the rest of the paper will be considered as galvanic. The Receiver (Rx) can either be capacitive or galvanic (Fig. \ref{fig:Intro}).

\subsection{Simulation Setup}
We perform the simulations using the Finite Element Method based EM solver Ansys High Frequency Structure Simulator (HFSS). A simplified model of a human body made of two crossed cylinders (Fig. \ref{fig:Sim_Setup} (a), (b), (c)) made up of skin and muscle tissue \cite{Gabriel_1996} is used to reduce computational complexity as well as time. The validity of this model has been shown by comparing the EM field distribution generated by the cross cylindrical model with a complex human body model - VHP Female v2.2 from Neva Electromagnetics \cite{neva_model} which showed identical results \cite{Safety_Study} . The simulations are carried out at 21 MHz, according to the IEEE 802.15.6 standard \cite{astrin2012ieee}.

The setup is shown in Fig. \ref{fig:Sim_Setup} with the structural parameters of the model. A galvanic Tx is used inside the body at a height of 80 cm from ground plane which can be placed in various orientations that are shown in Fig. \ref{fig:Sim_Setup} (d). The Rx can either be a galvanic or a capacitive Rx as illustrated in Fig. \ref{fig:Sim_Setup} (e).

\subsection{Simulation Results}

Fig: \ref{fig:Sim_Results} illustrates how the variation in orientation of the Tx changes the received voltage across the torso of the human body. For orientation 1 (Fig. \ref{fig:Sim_Results} (a)), we observe that when the Rx is placed close to the Tx (separation < 20 cm) but not at the same height as the Tx, a galvanic Rx has a lower channel loss than a capacitive Rx due to dipole coupling dominating the channel loss at smaller distances. Further, a null point is observed when the galvanic Rx is at the same height as that of the Tx due to the fields from two plates cancelling each other symmetrically out at that point. As for a capacitive Rx, we see that inter-device coupling results in channel loss reducing as the Rx is brought closer to the Tx. As the Rx moves away from the Tx, the received voltage saturates thus resulting in a saturation of the channel loss. It can be further observed that for larger distances between a Tx and Rx, the capacitive Rx has a higher received voltage than a galvanic Rx. \par
In orientation 2 (Fig. \ref{fig:Sim_Results} (b)), the two Tx plates are positioned in a manner that the field lines cancel each other out at the Rx end which results in the received voltage for both capacitive and galvanic being lower than that for orientation 1 and 3. 
Orientation 3 (Fig. \ref{fig:Sim_Results} (c)) results in the fields propagating outwards from the body towards the Rx reducing the path loss. Path loss from galvanic Rx keeps increasing with distance while capacitive Rx picks up the body potential which saturates as we move away from the Tx. However,the effect of two plates of the Tx canceling each other out occurs when Rx and Tx are at the same height resulting in lower received voltage.

\begin{figure}[b!]
\vspace{-1em}
\centering
\includegraphics[width=0.5\textwidth]{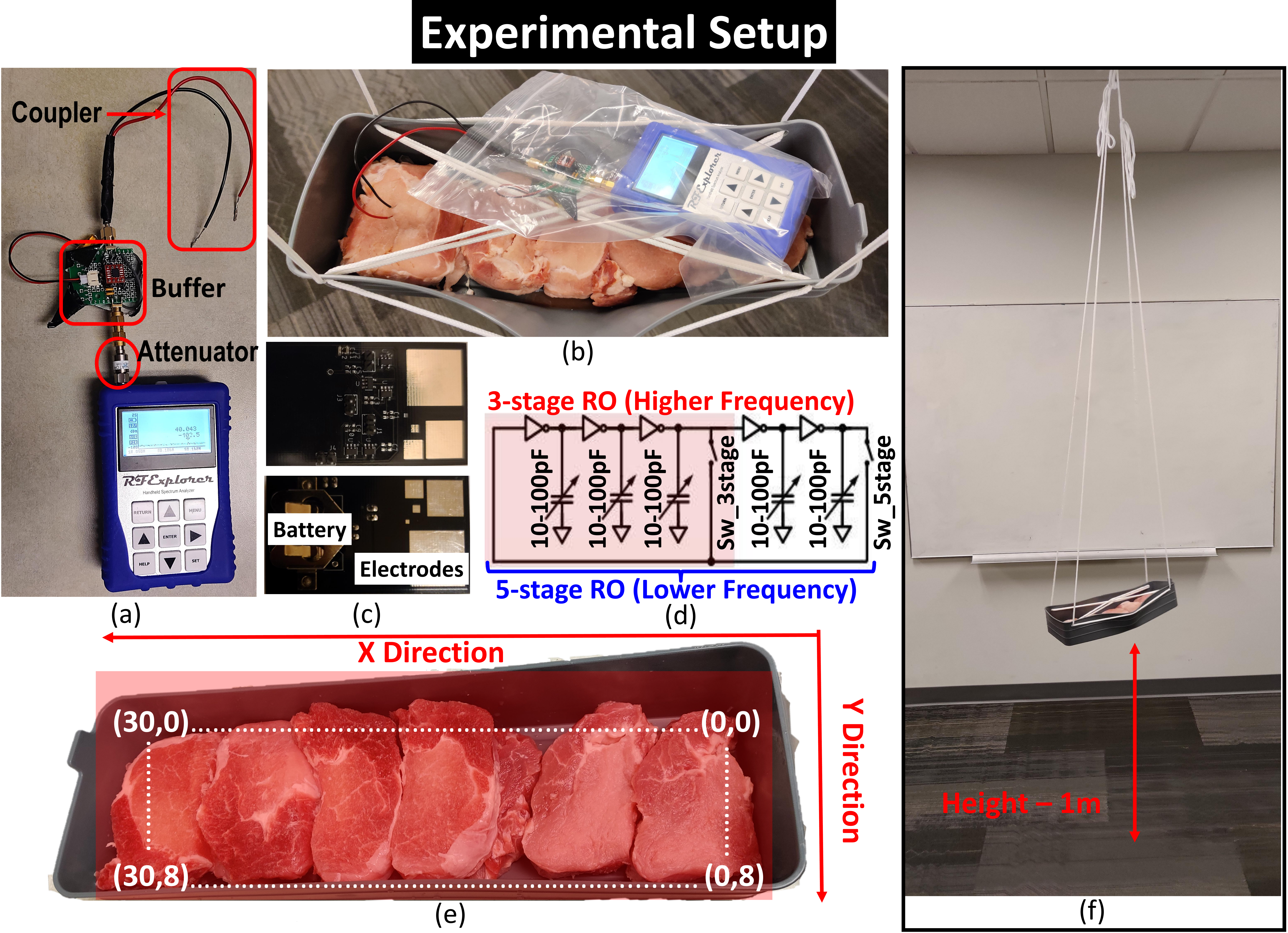}
\caption{(a) Rx setup with handheld spectrum analyzer, 20 dB attenuator and a high impedance buffer. (b) Rx electrodes touching the top surface of the meat to get the results. (c) Top and bottom surface of Tx node. (d) Ring oscillator circuit used for the Tx node. (e) Coordinates used to define the position of Rx for experiments. (f) Complete hanging box setup used for the experiments.}
\vspace{-1em}
\label{fig:Expt_Setup}
\end{figure}

\subsection{Experimental Setup}

The experiments were performed in a standard laboratory environment. The Rx setup (Fig. \ref{fig:Expt_Setup} (a), (b)) consists of a handheld RF Explorer spectrum analyzer. The use of a wearable spectrum analyzer is essential because wearable devices have a small ground plane with a smaller return path capacitance providing us with realistic results. Using a large tabletop spectrum analyzer will provide an optimistic path loss due to its large ground size. A broadband buffer - BUF602ID from Texas Instruments is used for a high impedance termination at the Rx end and a 20 dB attenuator is used to accurately pick up the high power signals where the wearable spectrum analyzer readings saturate. The receiver has an SMA connected coupler with two electrodes. For galvanic Rx, both electrodes touch the surface of the meat whereas for capacitive Rx, one of the electrodes is kept floating. The Galvanic Tx node (Fig. \ref{fig:Expt_Setup} (c)) is designed as a configurable 3-stage/5-stage ring oscillator (Fig. \ref{fig:Expt_Setup} (d)) on a 5.9cm $\times$ 3.6cm FR4 PCB using 74LVC1G04 inverters \cite{INV}.

An operating frequency of $\approx$21 MHz  is used as per the IEEE 802.15.6 standard. The Tx is placed in a plastic box within a layer of pork meat which is used to model the human tissue due to their close resemblance in dielectric properties. The experiment is carried out by taking measurements across the surface of the meat layer as illustrated in Fig. \ref{fig:Expt_Setup} (e). The Tx is positioned at the coordinate $(0,0)$ between two layers of meat while the Rx is moved across the "X-Y" plane on the surface of the meat. The meat box setup is kept hanging from the ceiling (Fig. \ref{fig:Expt_Setup} (f)) which is essential in accurately modeling the Rx's return path capacitance. Experiments performed on a tabletop provide optimistic path loss values when a capacitive Rx is used as higher return path capacitance occurs from a strong ground coupling generating inaccurate results.

\begin{figure}[t!]
\vspace{-1em}
\centering
\includegraphics[width=0.5\textwidth]{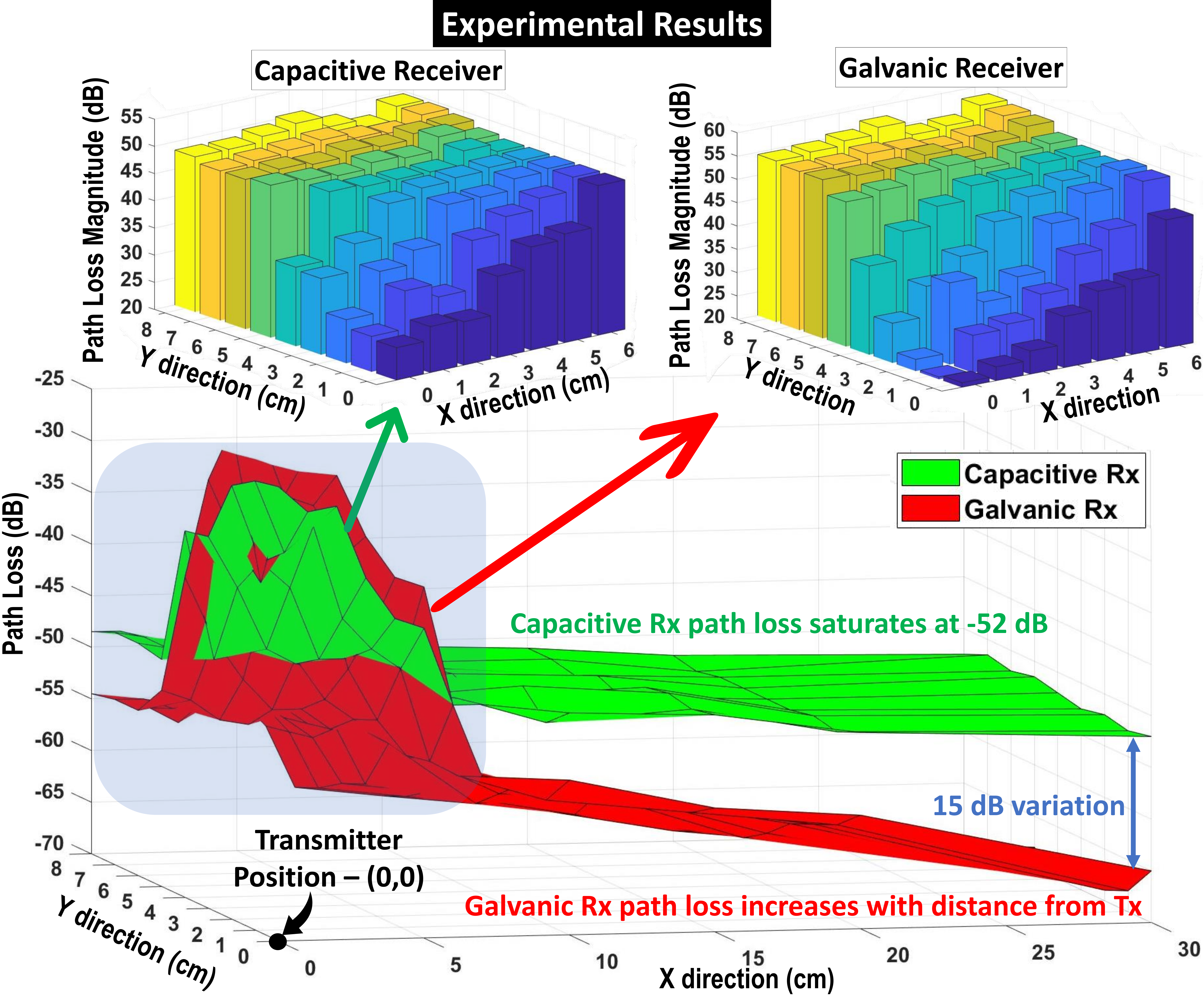}
\vspace{-1em}
\caption{Surface plot showing the path loss at different positions across the surface of the meat. The path loss magnitudes for shorter separations between Tx and Rx is plotted as 3D bar plots for capacitive and galvanic Rx. }
\vspace{-1.5em}
\label{fig:Expt_Results}
\end{figure}

\subsection{Results}

Experimental results plotted in Fig. \ref{fig:Expt_Results}  capture the trends shown in the simulations. The surface plot shows that the path loss for a capacitive Rx saturates at $-52 dB$ when Rx and Tx are placed away from each other (separation > 10 cm). For galvanic Rx, the path loss keeps increasing with distance between the Tx and the Rx. Further, when we move close to the Tx, path loss for both capacitive and galvanic Rx decreases. However, for separation of less than $5 cm$ between the Tx and Rx, we see that the galvanic Rx has lower path loss than a capacitive Rx. This has also been observed in the simulations (Fig. \ref{fig:Sim_Results} (a)). The path loss magnitude for capacitive and galvanic Rx has been shown in separate bar plots for the region shaded blue in the surface plot.

\section{Discussion on Receiver Design}
\begin{figure}[t!]
\centering
\includegraphics[width=0.35\textwidth]{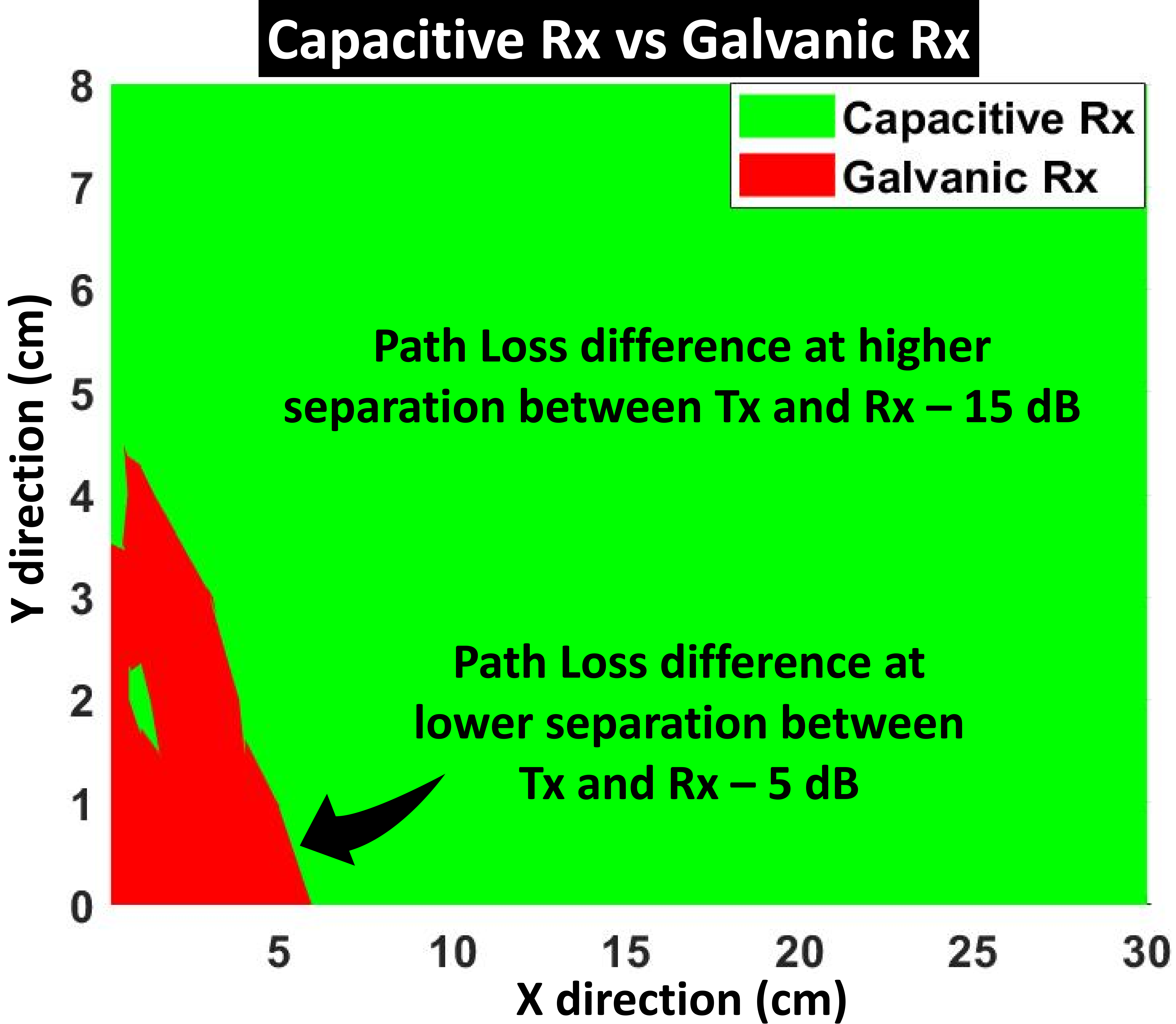}
\vspace{-0.5em}
\caption{Contour plot showing the regions where path loss for capacitive Rx is better versus where path loss for galvanic Rx is better.}
\vspace{-1.5em}
\label{fig:Expts_Contour}
\end{figure}

The path loss of the in-body to out-of-body communication system varies with the choice of the Rx used and the orientation of the Tx as illustrated by the results. Fig. \ref{fig:Expts_Contour} shows the regions where galvanic and capacitive Rx provide a lower path loss. The red part of the contour shows the region where galvanic Rx has a lower path loss when the Tx and Rx are close to each other. The green region is where capacitive Rx provides lower path loss when Tx and Rx separation is higher. \par

Hence, for smaller distances between Tx and Rx, a galvanic Rx can allow us to transmit at a lower power thus increasing battery life of the device. On moving away from the Tx, a capacitive Rx will have lower path loss response. Thus a reconfigurable Rx which changes between capacitive and galvanic modes as per the distance and orientation of the Tx can be used if the application focuses on small-distance communication to optimize the path loss response. However, the difference in path loss between capacitive and galvanic Rx for smaller separations is around $5 dB$ whereas the difference increases to $15 dB$ when the Rx is moved further away from Tx. If a high sensitivity receiver is being used, a capacitive Rx can be used across the whole body trading off $5 dB$ higher loss at smaller separations to reduce the design complexity.

\section{Conclusion}

The channel model for a physically secure Electro-Quasistatic Human Body Communication based system for in-body to out-of-body communication has been studied. The use of capacitive and galvanic receiver with a galvanic transmitter has been analyzed using finite element method based simulations and experiments. A reconfigurable receiver has been proposed which can convert between capacitive and galvanic modalities as per the separation between the transmitter and receiver as well as the orientation of the devices. This setup can be used to minimize the path loss between the devices for varying positions all across the body. Further, the trade-offs present in using a capacitive device as the only receiver have also been analyzed. 

\section*{Acknowledgment}
This work was supported by the National Science Foundation (NSF) Career Award under Grant CCSS 1944602. The authors wish to acknowledge the support of the USDA, through the NRI and CPS programs, under grants  2018-67007-28439 and 2019-67021-28990. The authors are with the School of Electrical and Computer Engineering, Purdue University, West Lafayette, IN 47907 USA. Corresponding author: Professor Shreyas Sen, e-mail: shreyas@purdue.edu 

\bibliographystyle{IEEEtran}

\bibliography{IEEE_example}

\end{document}